# Document Clustering using K-Means and K-Medoids

Rakesh Chandra Balabantaray*, Chandrali Sarma**, Monica Jha***

**Abstract**

With the huge upsurge of information in day-to-day's life, it has become difficult to assemble relevant information in nick of time. But people, always are in dearth of time, they need everything quick. Hence clustering was introduced to gather the relevant information in a cluster. There are several algorithms for clustering information out of which in this paper, we accomplish K-means and K-Medoids clustering algorithm and a comparison is carried out to find which algorithm is best for clustering. On the best clusters formed, document summarization is executed based on sentence weight to focus on key point of the whole document, which makes it easier for people to ascertain the information they want and thus read only those documents which is relevant in their point of view.

## 1. Introduction

Achievement of better efficiency in retrieval of relevant information from an explosive collection of data is challenging. In this context, a process called document clustering can be used for easier information access. The goal of document clustering is to discover the natural grouping(s) of a set of patterns, points, objects or documents. Objects that are in the same cluster are similar among themselves and dissimilar to the objects belonging to other clusters. The purpose of document clustering is to meet human interests in information searching and understanding. The challenging problems of document clustering are big volume, high dimensionality and complex semantics. Our motive in the present paper is to extract particular domain of work from a huge collection of documents using K-Means and K-Medoids clustering algorithm and to obtain best clusters which later on can be used for document summarizations. Document clustering is a more specific technique for document organization, automatic topic *extraction and fast* IR[1], which has been carried out using K-means clustering algorithm implemented on a tool named WEKA[2](Waikato Environment for Knowledge Analysis) and K-Medoids algorithm is executed on Java NetBeans IDE to obtain clusters. This allows a collection of various documents from different domains to be segregated into groups of similar domain. The best cluster obtained, then undergoes summarization to retrieve highest weighed sentences to reveal gist of the whole document.

## 2. Reviews

Throughout these years, there has been a lot of work implemented on document clustering using k-means by various researchers employing different means. Initially, the researchers worked using the simple k-means algorithm and then in later years, various modifications were executed. Use of K-mean clustering and Vector Space model was employed by using the text data by treating it as high dimensional. It was shown that the time taken for entire clustering process was linear in the size of document collection [Indrajit S. Dhillon et.al. 2001]. Some researchers found an effective technique for K-means clustering which proves that principal components are the continuous solutions to the discrete cluster membership [Chris Ding et.al.2004]. Recently, work on the performance of the partition clustering techniques in terms of complex data objects and comparative study of the cluster algorithm for corresponding data and proximity measure for specific objective function based on K-means and EM Algorithms was executed. Comparison and evaluation clustering algorithms with multiple data sets, like text, business, and stock market data was performed. Comparative study of clustering algorithms identified one or more problematic factors such as high dimensionality, efficiency, scalability with data size, sensitivity to noise in the data. [Satheelaxmi.G et.al.2012]. Work by integrating the constraints into the trace formulation of the sum of square Euclidean distance function of K-mean$s$ by combining criterion function transformation into trace maximization was optimized by eigen-decomposition [Guobiao Hu et.al.2008]. Plentiful times work on K-means algorithm on WEKA model had been implemented in the past which in turn has improved the WEKA tool set. [Sapna Jain et al.2010]. Previously, work based on several datasets, including synthetic and real data, show that the proposed algorithm may reduce the number of distance calculations by a factor of more than a thousand times when compared to existing algorithms while producing clusters of comparable quality was carried out[Maria Camila N. et.al.2006]. Proposal of a new language model, to simultaneously cluster and sum up at the same time has been implemented in past. The method implies good document clustering method with more meaningful interpretation and a better document summarization method taking the document context information into consideration [Wang, Shenghuo Zhu et al.2008].

\* IIIT Bhubaneswar, Bhubaneswar, Odisha, India. E-mail: rakeshbray@gmail.com
\*\* Department of Information and Technology, Gauhati University, Guwahati, India. E-mail: chandralis.88@gmail.com
\*\*\* Department of Information and Technology, Gauhati University, Guwahati, India. E-mail: monica.jha@rediffmail.com

## 3. Methodology

Execution of algorithm is based on the input provided to the algorithm. In this paper, the input provided to the algorithm had to undergo certain refinement. The experiment generally comprises of four major steps.

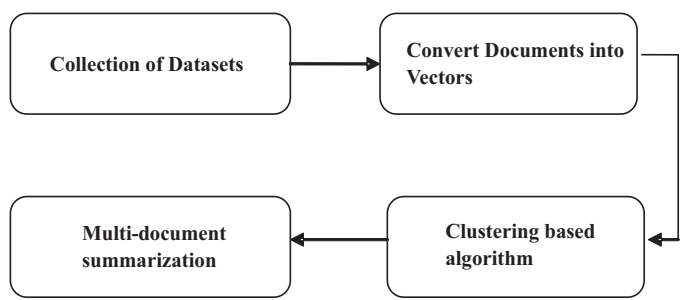

Same input is provided to both the algorithms and later on after the algorithm implementation is over, the best cluster obtained is then used for document summarization.

### 3.1. Collection of Datasets

At the beginning hundred documents were collected which consisted twenty each of Entertainment ($e1, e2….e20$), Literature ($l1, l2….l20$), Sport ($s1, s2….s20$), Political ($p1, p2…. p20$) and Zoology ($z1,z2,….z20$). These documents undergo refinement which is fed to the algorithm to obtain clusters containing documents from similar domains.

### 3.2. Convert Documents into Vectors

A document usually consists of huge number of words, it is not always necessary that each word is of importance. Due to which, a document has high dimensionality has to be reduced. Hence processing is carried on a document to reduce this dimensionality and get rid of extra words and to obtain weight of each of the word to be used in the algorithm. Conversion of documents into vectors is carried in various steps.

### 3.2.1. Tokenization

All processes in information retrieval require the words of the data set. The main use of tokenization is identifying the meaningful keywords called tokens. Tokenization splits sentences into individual tokens, typically words.

### 3.2.2. Stop-Words Removal

Collected documents contain some unnecessary words by which dimensionality of a document will be increased; we should remove those words to get proper result. Pronoun, adverb, preposition etc. which are used constantly throughout in a document has to be removed.

### 3.2.3. Weight Calculation

This step involves calculating the weight of each word twice, once using frequency of words and then using term frequency-inverse document frequency (tf-idf).

a. Weight calculated using frequency

Ratio of each word occurred in the document to the total number of words in that document gives us weight.

Weight = (frequency) ÷ (total number of words in a document)

b. Weight calculated using tf-idf method

The other method which is often used in information retrieval and text mining known as tf-idf weight, where tf is the "Term Frequency" and is denoted $tf_{t,d}$ with the subscripts denoting the term and the document in order and idf is the "Inverse Term Frequency". The term vector for a string is defined by its term frequencies. If count (t,cs) is the count of term t in character sequence cs, then the term frequency (TF) is defined by,

$$Tf(t,cs) = sqrt(count(t,cs))$$

If df(t,docs) be the document frequency of token t i.e. the number of documents in which the token t appears, then the inverse term frequency (IDF) of t is defined by,

$$idf(t,docs) = log(D/df(t,docs))$$

Therefore if term weight is denoted by Wi then typically,

$$W_i = TF * IDF$$

### 3.2.4. Weighted Matrix Formation

A matrix had to be formed where we stored weight of each word in rows and in the column we had stored documents name. This matrix serves as the input for both the algorithm. For formation of this matrix we employ the idea from Vector Space Model (VSM)[3] which is an algebraic model for representing text documents (and any objects, in general) as vectors of identifiers, such as, for example, index term.

$$\begin{array}{c} & Tw1 & Tw2 & Tw3 & Tw4 & Tw5 \\ d1 \\ d2 \\ d3 \\ d4 \end{array} \begin{pmatrix} 0.5 & 0.25 & 0.25 & 0.125 & 0.0 \\ 0.15 & 0.234 & 0.44 & 0.67 & 0.56 \\ 0.55 & 0.56 & 0.23 & 0.25 & 0.125 \\ 0.45 & 0.26 & 0.5 & 0.55 & 0.0 \end{pmatrix}$$

The matrix formed has been used as input for both the algorithm.

### 3.3. Clustering Based Algorithms

Among various clustering based algorithm, we have selected K-means and K-Medoids algorithm. Implementation of K-means algorithm was carried out via WEKA tool and K-Medoids on java platform. Using the same input matrix both the algorithms is implemented and the results obtained are compared to get the best cluster.

K-Mean Clustering using WEKA Tool

To cluster documents, after doing preprocessing tasks we have to form a flat file which is compatible with WEKA tool and then send that file through this tool to form clusters for those documents. This section will give a brief mechanism with WEKA tool and use of K-means algorithm on that tool. WEKA has four applications Explorer, Experiment, Knowledge Flow and Simple CLI but here only Explorer application and the two fields from this application i.e. preprocess and cluster fields are used.

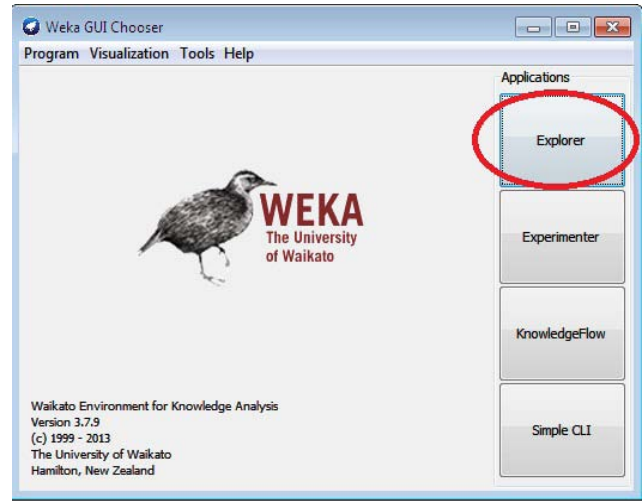

### 3.3.1. WEKA Data File Format (input)

As it is already mentioned that weighted matrix is the input for the implementation of the algorithm, so an .arff file is formed (which is the common file extension for WEKA) which consists of the matrix formed earlier. Two types of attributes are used for this experiment, one is the numerical attribute which represents weights of each term in the matrix and other is the nominal attribute which represents list of documents to be clustered. Since the row of the matrix contains list of documents so nominal attribute represents rows and column contains 450 terms so there are 450 numeric attributes (w1 to w450) present in the file.

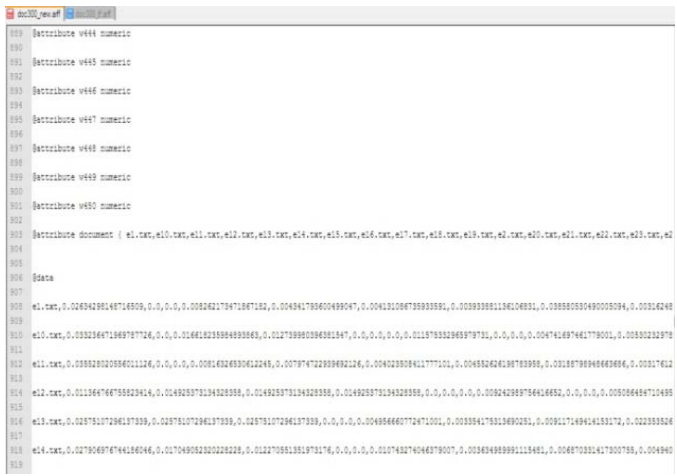

### 3.3.2. Load Data

To apply this tool first we have to load the input file to the tool. To load data, first the "Open File" button should be clicked and after clicking this button the .arff file is loaded as shown in the figure. After loading the file and click on "Open" button it will display a list of attributes in the window which contains number and name of attributes.

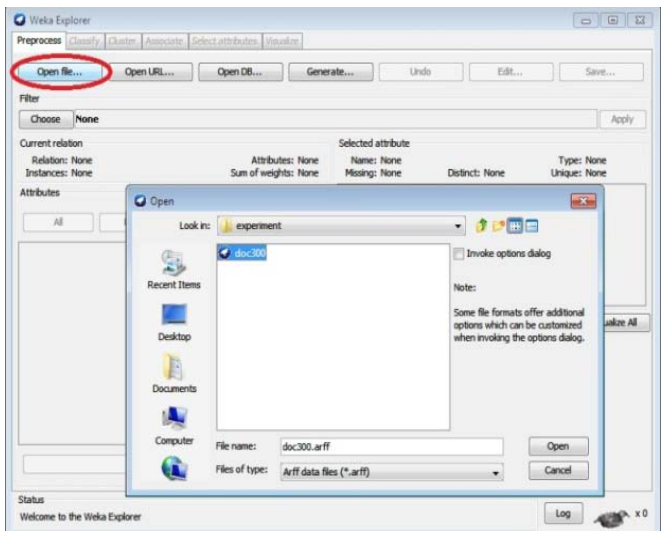

### 3.3.3. Cluster Data using K-means

The next field which is used in this experiment is Cluster field. To perform clustering, we have to select the "Cluster" tab in the Explorer and click on the "Choose" button. This results in a drop down list of available clustering algorithms. From those algorithms "SimpleKmeans" is selected for our experiment.

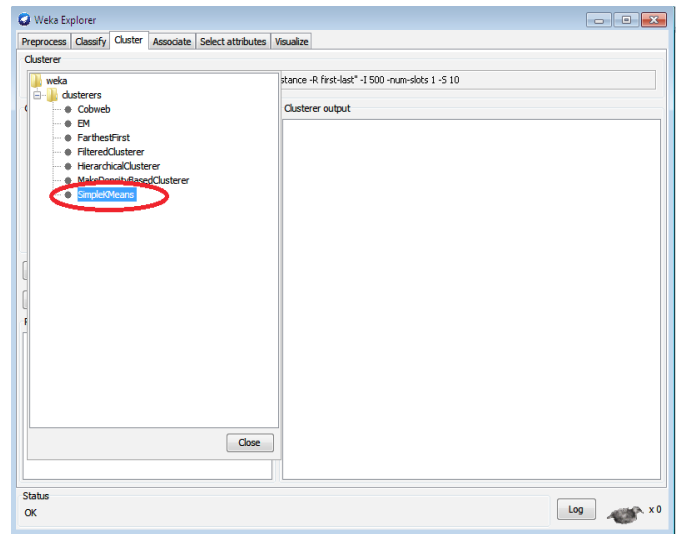

To obtain the cluster, clustering parameter (number of clusters and distance functions) had to be changed. For our experiment number of cluster is set to 5 and clustering is done for both Euclidian distance[4] and Manhattan distance[5] functions.

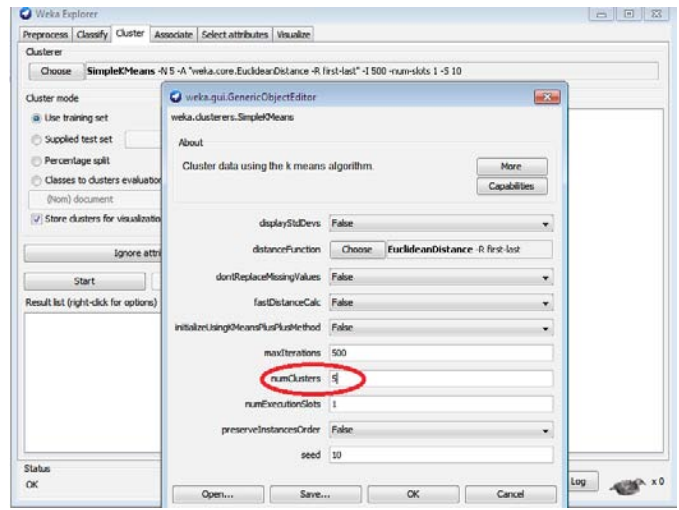

Now clicking "Start" button will give a result set on the right side of the window. We can right click the result set in the "Result list" panel and view the results of clustering in a separate window. This resulting window is shown in the figure below.

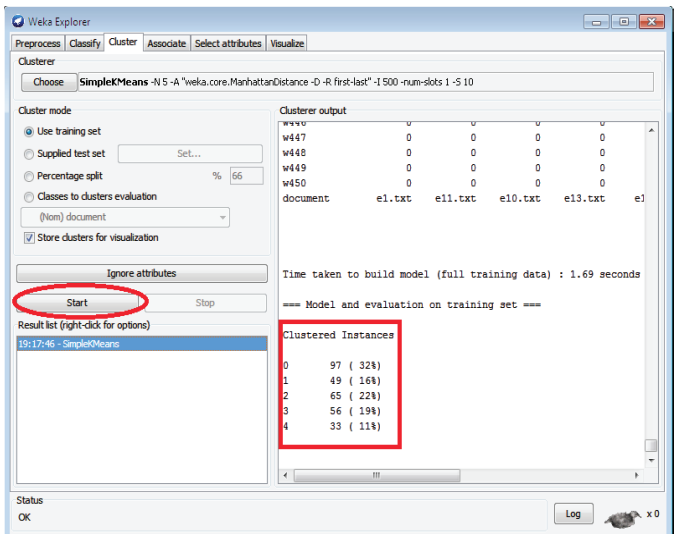

WEKA tool has an ability to visualize clusters which are already formed. To visualize those clusters in a 2D graph we can right click the result set in the "Result list" panel and click on "Visualize cluster assignments" to visualize it in a separate window. In 2D graph clusters number (cluster0, cluster1, cluster2, cluster3 and cluster4) is chosen as X-axis and name of the documents (e1…e60, l1….l60, p1….p60, s1…s60, z1…z60) as Y-axis. This will give us the final cluster formation for K-means clustering algorithm.

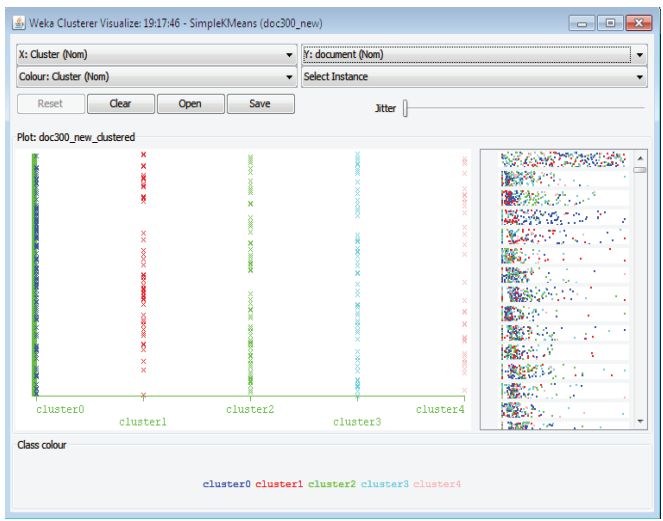

K-Medoids Clustering Algorithm

Clustering based algorithm which was developed to improve the result obtained by K-means algorithm as the later is sensitive to outliers hence does not produce satisfactory results, which is not same in the case of K-Medoids algorithm. K-Medoids algorithm is partition based. Steps followed in the execution of the algorithm:

### 3.3.4. Algorithm

*Input:* k – the number of clusters to be partitioned.

*Output:* A set of '*k*' clusters that minimizes the sum of the dissimilarities of all the objects to their nearest medoid.

   n – The number of objects.
- Arbitrarily choose 'k' objects as the initial medoids;
- Repeat,
  a. Assign each remaining object to the cluster with the nearest medoid;
  b. Randomly select a non-medoid object;
  c. Compute the total cost of swapping old medoid object with newly selected non-medoid object.
  d. If the total cost of swapping is less than zero, then perform that swap operation to form the new set of k- medoids**.**
- Until no change.

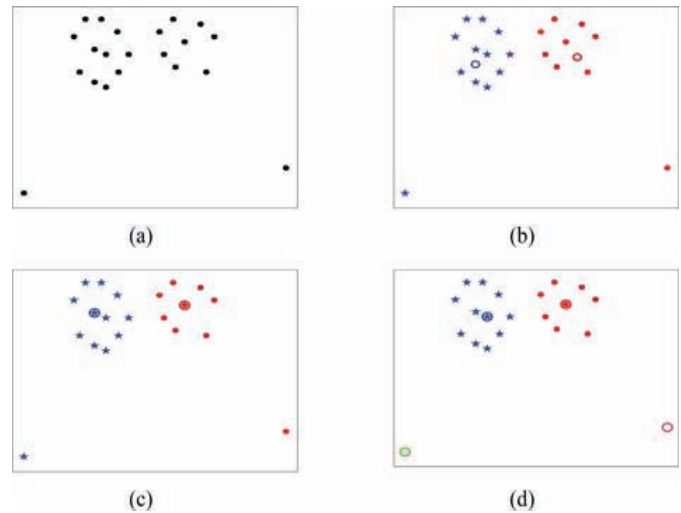

Different Steps of K-Medoids Algorithm

It was mentioned earlier that a matrix of documents and term weight is formed to use as input in the algorithm. A matrix of hundred documents was formed during pre-processing and then a program was constructed on java platform on which the matrix was fed which at the end yielded clusters. At the end, there was comparison made of the clusters formed using both the algorithms to find in the best cluster which is later utilized for document summarization to reveal the key-points of each of the documents of the best cluster formed based on the weight of each word of the document.

## 4. Experimental Results

Implementing both the algorithms, certain clusters are obtained. Beginning of the implementation was carried out by taking hundred documents from following domains:

1. Entertainment (20 documents)
2. Literature (20 documents)
3. Political (20 documents)
4. Sport (20 documents)
5. Zoology (20 documents)

Two varied result was obtained after applying each algorithm once on the weight calculated using term frequency and the other using tf-idf method.

### 4.1. Result for K-meansmanh

Better result was obtained using tf-idf weight, that result has been considered as the final cluster result for K-means clustering.

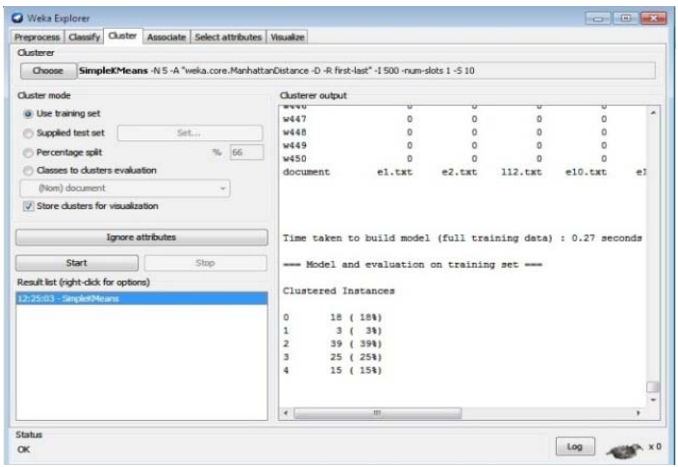

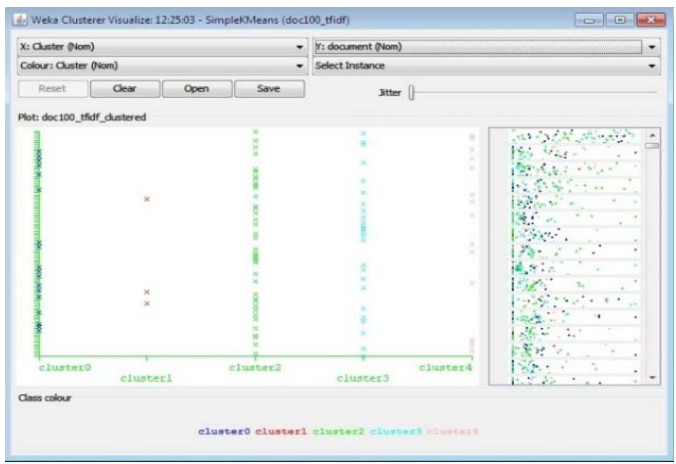

### 4.1.1. Result for K-Medoids

The result obtained after implementing the algorithm using tf-idf weight has been considered for comparison.

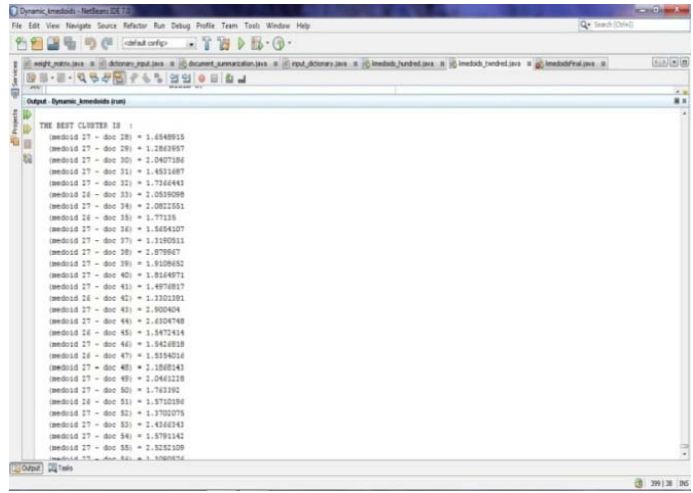

### Comparison of the Results

A dataset comprising of hundred documents of five different domains was taken as input and was used to obtain clusters by executing K-means and K-Medoids algorithms. Then a comparison of the best clusters formed from both the algorithm was made to incur the best algorithm for clustering.

| Algorithm | Comparison | | |
|---|---|---|---|
| | Cluster number | Number of documents | Efficiency |
| K-Means | 0 | 18 | 27.78% |
| | 1 | 3 | 66.67% |
| | 2 | 39 | 25.64% |
| | 3 | 25 | 24% |
| | 4 | 15 | 26.67% |
| K-Medoids | 0 | 43 | 18.60% |
| | 1 | 11 | 54.15% |
| | 2 | 7 | 42.85% |
| | 3 | 35 | 14.2% |
| | 4 | 4 | 50% |

In the above table, each cluster number defines documents of a particular domain. From the above result for hundred documents we determine that the clusters obtained using K-means algorithm is more efficient than the clusters obtained from K-Medoids algorithm.

The best result obtained then undergoes summarization to display the key points of each documents based on weight of each sentence of a document. Because of this, it becomes easier for the reader to retrieve the most relevant document according to his requisite.

## 5. Conclusion

The paper analyzes two clustering algorithm K-means and K-Medoids to obtain best cluster using the refinement of input to the algorithm, employed. Both the algorithms were executed on a dataset of hundred documents for clustering. K means algorithm was carried out using both Euclidean and manhattan distance on WEKA tool and K-medoids was carried out through java programming. After the completion of execution of both the algorithms, it was observed that k-means yields better result than k-medoids. The clusters formed in k-means algorithm is more efficient than that of the k-medoids, as because manhattan distance of k-means obtained is better than the Euclidean distance obtained also non- normalized tf-idf gives better result. Moreover, huge document size also gives an advantage to the K-means method because this improves the similarity measure. Hence, it has been ascertained in the end that the best cluster is obtained using K-means algorithm which can later be used for multi-document summarization. The summarization helps the user in saving their time by providing just the key points of a document which represents the category of the document because of which the user can easily retrieve the document they want to access.

## 6. Acknowledgement

We would like to thank the college authorities of IIIT Bhubaneswar for providing us with a suitable research environment necessary for successful completion of our work.

## References

1. Dhillon, I. S., Fan, J. & Guan, Y. (2001). Efficient Clustering of Very Large Document Collections (Chapter 1). doi:10.1145/502512.502550
2. Ding, C. & He, X. (2004). *K-means Clustering via Principal Component Analysis,* 225-232.
3. Satheelaxmi, G., Murty, M. R., Murty, J. V. R. & Reddy, P. (2012). Cluster analysis on complex structured and high dimensional data objects using K-means and EM algorithm. *International Journal of Emerging Trends & Technology in Computer Science*, 1(1).
4. Hu, G., Zhou, S., Guan, J. & Hu, X. (2008). Towards effective document clustering: A constrained K-means based approach. *Information, Processing and Management*, 44(4), 1397-1409.
5. Jain, S., Aalam, M. A. & Doja, M. N. (2010). *K-means Clustering Using Weka Interface*. Proceedings of the 4[th] National Conference; INDIACom-2010. New Delhi: Bharati Vidyapeeth's Institute of Computer Applications and Management.
6. Barioni, M. C. N., Razente, H. L., Traina, A. J. M. & Traina, C. Jr. (2006). An Efficient Approach to Scale Up K-medoid Based Algorithms in Large Databases.
7. Wang, D., Zhu, S., Li, T., Chi, Y. & Gong, Y. (2008*). Integrating Clustering and Multi-Document Summarization to Improve Document Understanding.*